\documentclass[rmp,twocolumn]{revtex4}
\usepackage[english]{babel}
\usepackage{amssymb,amsfonts,amsmath}
\usepackage{color}
\usepackage{graphicx}
\usepackage{amsmath}
\usepackage{natbib}
\usepackage{hyperref}
\usepackage{enumerate}

\usepackage{listings}
\newcommand{\EQ}[1]{Eq.~(\ref{eq:#1})}

\newcommand{\FIG}[1]{Fig.~\ref{fig:#1}}

\newcommand{\mut}{u}

\newcommand{\mexpfit}{\langle e^{F}\rangle}
\newcommand{\ox}{r}

\newcommand{\gt}{g}
\newcommand{\locus}{s}
\newcommand{\locuspm}{t}
\newcommand{\OO}{\mathcal{O}}

\definecolor{dkgreen}{rgb}{0,0.6,0}
\definecolor{gray}{rgb}{0.5,0.5,0.5}
\definecolor{mauve}{rgb}{0.58,0,0.42}

\begin{document}
\title{FFPopSim: An efficient forward simulation package for the evolution of
large populations}
\author{Fabio~Zanini}
\author{Richard~A.~Neher}
\affiliation{Max Planck Institute for Developmental Biology, 72076 T\"ubingen, Germany}

\date{\today}

\begin{abstract}
The analysis of the evolutionary dynamics of a population
with many polymorphic loci is challenging since a large number of
possible genotypes needs to be tracked. In the absence of analytical solutions, 
forward computer simulations are an important tool in
multi-locus population genetics. The run time of standard algorithms to simulate
sexual populations increases as $8^L$ with the number $L$ of loci, or with the
square of the population size $N$. 
We have developed algorithms that allow to
simulate large populations with a run-time that scales as $3^L$. The algorithm
is based on an analog of the Fast-Fourier Transform (FFT) and allows for
arbitrary fitness functions (i.e.~any epistasis) and genetic maps.
The algorithm is implemented as a collection of C++ classes and a Python
interface. {\bf Availability:} \url{http://code.google.com/p/ffpopsim/}.
\end{abstract}
\maketitle
\section*{Introduction}
Forward simulations of population genetics, track either 
the genotype of every individual in the population, or the number of
individuals that carry a particular genotype. The former has been implemented in
a number of very flexible simulation packages
\citep{Guillaume:2006p47092,Spencer:2004p47081,Peng:2005p47083}. In large
populations with a moderate number of loci, storing the abundance of all
possible $2^L$ genotypes is often faster. Simulating such large populations with
a small number of loci is for example essential when studying the
evolution of drug resistance in viral or bacterial pathogens
\citep{Weinreich:2006p15529}.

Individual-based population genetic simulations are quite straightforward and
usually employ a discrete generation scheme in which processes such as mutation,
selection, and migration are applied at every generation to every individual.
Individuals are then paired up via a mating scheme and recombinant offspring is
produced.
Existing toolboxes often emphasize biological realism and allow the user to
specify complex life cycles, see e.g.~\citet{Guillaume:2006p47092}. Our emphasis
here is on efficient simulation of large populations.
Instead of tracking individuals, we keep track of the distribution $P(\gt)$ of
gametes across all possible $2^L$ genotypes, denoted by $\gt = ( \locus_1,
\ldots , \locus_L)$ where $\locus_i = 0/1$. This genotype distribution changes due to mutation,
selection and recombination. The former two are again straightforward and
require at most $L\,2^{L}$ operations (each genotype can mutate at any one of
the $L$ loci). In our implementation, selection acts on haploid gametes,
precluding dominance effects. Recombination, however, is a computationally
expensive operation since it involves pairs of parents (of which there are
$4^L$) which can combine their genome in many different ways ($2^L$). As a
consequence, a naive implementation requires $8^L$ operations to calculate the
distribution of recombinant genotypes.
It is intuitive that the complexity of this algorithm can be reduced:
given a recombination pattern, only a fraction of the genome is passed on in
sexual reproduction and all genotypes that agree on that fraction contribute
identically.  We will show below that exploiting this redundancy allows to
reduce the number of operations from $8^L$ to $\mathcal{O}(3^L)$. 

After selection, mutation, and recombination, the
population distribution $P(\gt)$ contains the expected number of individuals
of genotype $\gt$ in the next generation. For stochastic population genetics, we
still need to resample the population in way that mimics the randomness of
reproduction. This is achieved by resampling individuals according to a Poisson
distribution with mean $NP(\gt)$ for each genotype, which will result in a
population size of approximately $N\pm \mathcal{O}(\sqrt{N})$. 

\section*{Recombination via FFT}
The probability of producing a genotype $\gt$ by recombination is
\begin{equation}
R(\gt) = \sum_{ \xi } \sum_{\gt'} C(\xi) P(\gt^m)P(\gt^p),
\end{equation}
where $ \xi := (\xi_1,\ldots,\xi_L)$ specifies the particular way the parental
genomes are combined:
$\xi_i=0~\text{(resp. $1$)}$ if locus $i$ is derived from the mother (resp.
father).
The genotype $\gt'$ is summed over; it represents the parts of the maternal
($\gt^{m}$) and paternal ($\gt^{p}$) genotypes that are not passed on to the
offspring.
We can decompose each parent into successful loci that made it into the
offspring and wasted loci, as follows: $\gt^{p} = \xi \land \gt + \bar{\xi}
\land \gt'$ and $\gt^{m} = \bar{\xi} \land \gt + \xi \land \gt'$, where $\land$
and a bar over a variable indicate respectively the elementwise AND and NOT
operators. The function $C$ assigns a probability to each inheritance pattern,
see \citet{Neher:2011p45096} for a more detailed explanation.
In a facultatively sexual population, a fraction $\ox$ of $P(\gt)$ is replaced
by $R(\gt)$,  while $\ox=1$ in an obligate sexual population.

The central ingredient for the efficient recombination is a fast-Fourier
algorithm for functions on the L-dimensional binary hypercube.
Every function on the hypercube can be expressed as
\begin{equation}
F(\gt) = f^{(0)} + \sum_{i} \locuspm_i f^{(1)}_i + \sum_{i<j} \locuspm_i
\locuspm_j f^{(2)}_{ij} + \cdots
\end{equation}
where $\locuspm_i := (2\locus_i-1)$ and takes the values $\pm 1$. 
In total, there are $2^L$ coefficients $f^{(k)}_{i_1\ldots i_k}$ for every
subset of $k$ loci out of a total of $L$ loci \citep{Weinberger:1991p17543}. Similarly, each
coefficient $f^{(k)}_{i_1\ldots i_k}$ is uniquely specified by
\begin{equation}
f^{(k)}_{i_1\ldots i_k} = 2^{-L}\sum_{\gt} \locuspm_{i_1}\ldots\locuspm_{i_k} F(\gt).
\end{equation}
These nominally $4^L$ operations can be done in $L\, 2^L$ via the FFT scheme
illustrated in \FIG{FFT}.

\begin{figure}[tb]
\begin{center}
  \includegraphics[width=0.98\columnwidth]{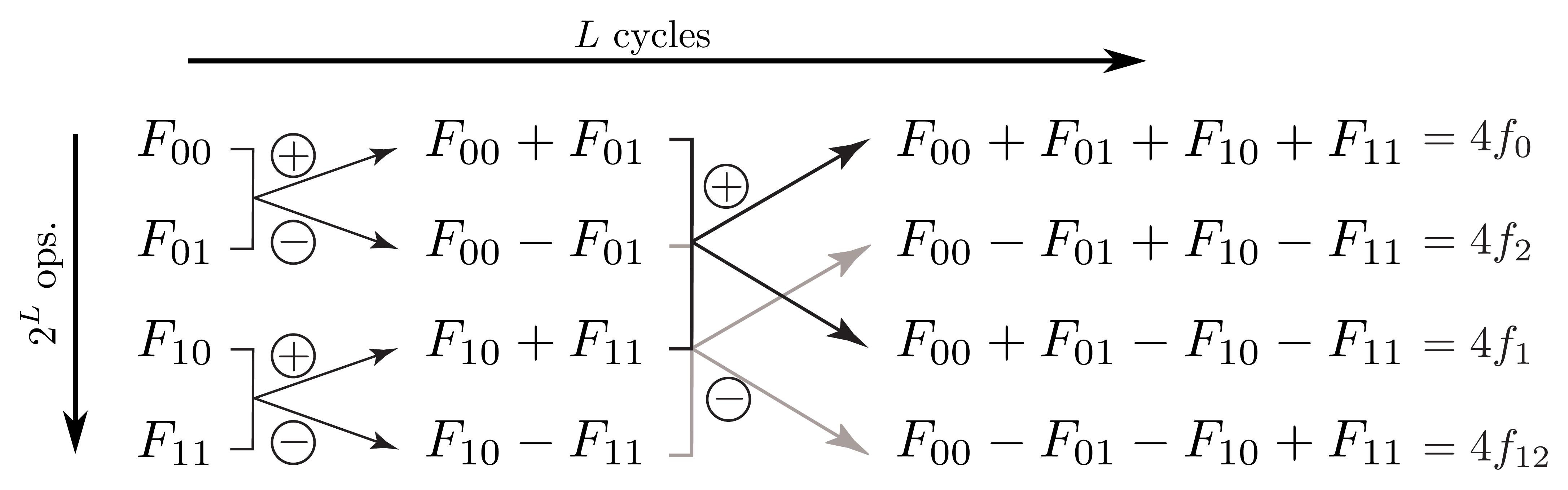}
  \caption[labelInTOC]{Calculating the Fourier transform in $L\, 2^L$ operations.
  Arrow going up indicate addition, going down substraction. For the general
  $L$ dimensional hypercubes $L$ cycles are necessary where terms differing at
  different bits are combined.}
  \label{fig:FFT}
\end{center}
\end{figure}

With some algebra (see online supplement), one can show that the generic
Fourier coefficient of $R(\gt)$ is given by
\begin{equation}
r^{(k)}_{i_1\ldots i_k} = \sum_{l=0}^k \sum_{j, \, h} C^{(k)}_{j_1\ldots
j_l h_1\ldots h_{k-l}} \, p^{(k)}_{j_1 \ldots j_l}\,  p^{(k-l)}_{h_1 \ldots h_{k-l}},
\end{equation}
where the sum runs over all partitions of $i_1\ldots i_k$ into groups of $l$ and
$k-l$ denoted by $j:=(j_1,\ldots,j_l)$ and $h:=(h_1,\ldots, h_{k-l})$. Variables
such as $p^{(k)}_{j_1 \ldots j_l}$ are the Fourier coefficients of the genotype distribution,
$P(g)$, and the crossover function $C$ is expanded into  
\begin{equation}
C^{(k)}_{j_1 \ldots j_l h_1 \ldots  h_{k-l}} := \sum_{\xi}
C(\xi) \, \xi_{j_1}\ldots\xi_{j_l}\bar{\xi}_{j_1}\ldots\bar{\xi}_{h_{k-l}}.
\end{equation}
The latter coefficients can be calculated efficiently by realizing that for
$k=L$, there is exactly one term unequal to zero. All subsequent terms can be
calculated by successive marginalization of unobserved loci. In total,
calculating $r^{(k)}_{i_1\ldots i_k}$ requires $\OO(2^k)$ operations. Since
there are ${L \choose k }$ terms of order $k$, the entire calculation requires
$\OO(3^L)$ operations. In case of single crossover recombination, the algorithm
can be sped up further to $\OO(L2^L)$.

\section*{Usage}
FFPopSim is implemented in C++ with a Python2 wrapper.
Documentation, a number of examples, and test routines are provided. As an
example, we discuss here the problem of fitness valley crossing, which has
received attention recently in the population genetics literature
\citep{Lynch:2010p36165,Weissman:2010p37077}. Consider a fitness landscape
where the wild-type genotype has (malthusian) fitness $s_1$, while the quadruple
mutant has fitness $s_1+s_2$. All intermediate genotypes have the same slightly
deleterious fitness $0$ ($-s_1$ relative to wild-type). The time required for
crossing the valley can be computed by the following routine:
{
\small
\begin{verbatim}
import FFPopSim
L = 4 # number of loci
N = 1e10 # population size
# create population and set rates
c = FFPopSim.haploid_lowd(L)
c.set_recombination_rates([0.01] * (L-1))
c.set_mutation_rate(1e-6)

# start with wildtype: 0b0000 = 0
c.set_genotypes([0b0000],[N])
# set positive relative fitness for wildtype
# and quadruple mutant: 0b1111 = 15
c.set_fitness_function([0b0000, 0b1111], 
                       [s1, s1+s2])
# evolve until the quadruple mutant spreads
while c.get_genotype_frequency(0b1111)<0.5: 
    c.evolve(100)
print c.generation
\end{verbatim}
}
The runtime and memory requirements of $3^L$ still preclude the simulation of
more than $L=20$ loci. For this reason, we also include a streamlined individual based
simulation package with the same interface that can simulate arbitrarily large
number of loci and has an overall runtime and memory requirements
$\mathcal{O}(NL)$ in the worst case scenario. To speed up the simulation in many
cases of interest, identical genotypes are grouped into clones. This part of the
library was developed for whole genome simulations of large HIV populations
($L=10^4$, $N>10^5$) and a specific wrapper class for HIV applications is provided.
As of now, the library does not support dominance effects which would require a
fitness function that depends on pairs of haploid genomes. Such an extension to
diploid populations is straightforward. 

\section*{Acknowledgement}
We would like to thank Boris Shraiman for many stimulating discussion and
pointing us at the FFT algorithm. This work is supported by the ERC though
Stg-260686.


\appendix\onecolumngrid
\section{Prerequisites and compiling}
The toolbox makes extensive use of the GNU scientific library
(\url{http://www.gnu.org/software/gsl/}) and the Boost C++ library
(\url{http://www.boost.org/}). The Python wrapper further requires NumPy, SciPy
and MatplotLib (\url{http://www.scipy.org}). If all of these are installed and
the appropriate path are set, FFPopSim can be compiled using Make.
Installation instructions are provided in the \texttt{INSTALL} file.
The building process creates files inside the folder \texttt{pkg}; C++ headers
are created in \texttt{pkg/include}, the static C++ library in \texttt{pkg/lib},
and the Python module files in \texttt{pkg/python}.

\section{Description of FFPopSim}
\texttt{FFPopSim} contains two packages of C++ classes and Python wrappers for
forward population genetic simulations.
One for large populations and relatively few loci ($L<20$), another one for
longer genomes. The former is called \texttt{hapoid\_lowd} and
tracks the abundance of all possible genotypes. The other one is called
\texttt{haploid\_highd} and tracks only genotypes present in the population. The
latter only allows for limited complextity of fitness functions and crossover
patterns. These two parts of the library have very similar syntax but work quite
differently under the hood. We will therefore describe them separately below.

A complete documentation in html is generated automatically from the source
using Doxygen and can be found in \texttt{doc/html/index.html}.

\subsection{FFPopSim for few loci}
\subsubsection{Specification of the population and the fitness function}
\lstset{language=C++, basicstyle=\small, showstringspaces=false, backgroundcolor=\color{white}, rulecolor=\color{black}, tabsize=2, keywordstyle=\color{mauve}, commentstyle=\color{dkgreen}, stringstyle=\color{blue} }
Since we assume that each locus is in one of two possible states $\locus_i=0/1$,
the genotype space is an $L$ dimensional binary hypercube. The population is a distribution
of individuals on that hypercube, and so are the mutant and recombinant
genotypes. Also the fitness function $F(\gt)$, which assigns a number to each genotype $\gt
= \{\locus_1,\ldots,\locus_L\}$, is a function on the hypercube. For
this reason, \texttt{hapoid\_lowd} makes extensive use of a class \texttt{hypercube\_lowd} that stores an
$L$ dimensional hypercube and implements a number of operations on the hypercube,
including a fast-Fourier transform (FFT). 

Every function on the hypercube can be expressed as
\begin{equation}
F(\gt) = f^{(0)} + \sum_{i} \locuspm_i f^{(1)}_i + \sum_{i<j} \locuspm_i
\locuspm_j f^{(2)}_{ij} + \cdots
\end{equation}
where $\locuspm_i := 2\locus_i-1$, i.e., simply a mapping from $\{0,1\}$ to
$\{-1, +1\}$.
There are $\binom{L}{k}$ coefficients $f^{(k)}_{i_1\ldots i_k}$ for every
subset of $k$ loci out of $L$ loci, so in total $2^L$ coefficients~\citep{Weinberger:1991p17543}. A
coefficient $f^{(k)}_{i_1\ldots i_k}$ is uniquely specified by
\begin{equation}
f^{(k)}_{i_1\ldots i_k} = 2^{-L}\sum_{\gt} \locuspm_{i_1}\ldots\locuspm_{i_k}
F(\gt)
\end{equation}
These nominally $4^L$ operations ($2^L$ for each coefficient) can be done in $L\, 2^L$ via the FFT scheme
illustrated in \FIG{FFT}. Both the forward and reverse transform are implemented
in \texttt{hypercube\_lowd}.
\begin{figure}[tb]
\begin{center}
  \includegraphics[width=0.78\columnwidth]{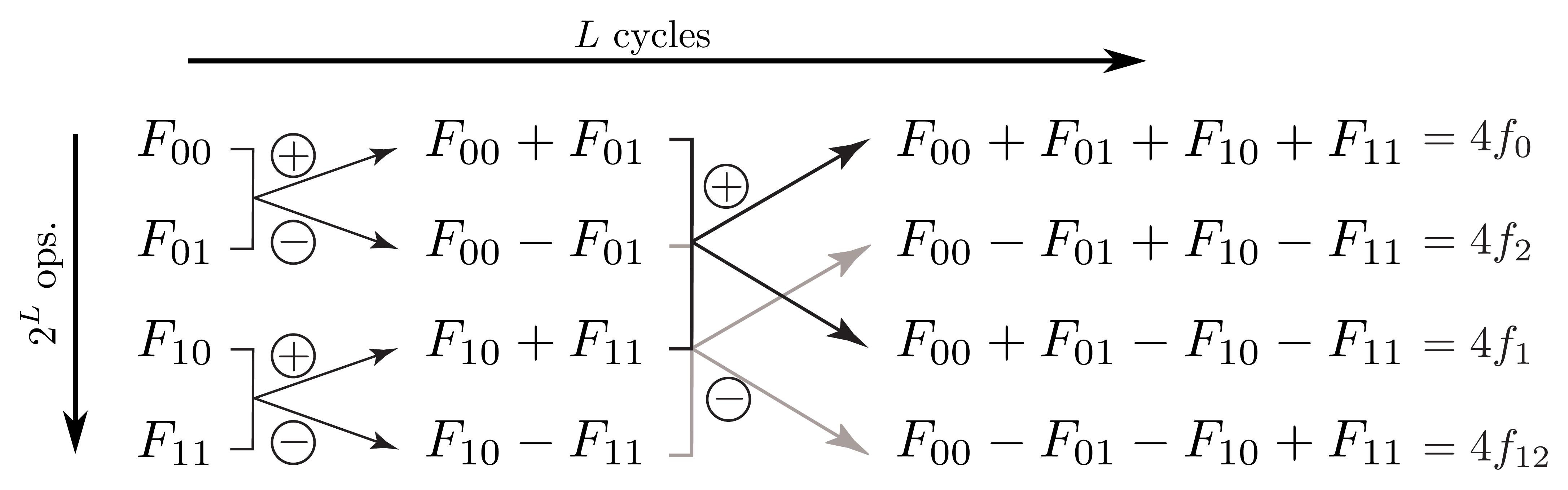}
  \caption[labelInTOC]{Calculating the Fourier transform in $L2^L$ operations.
  Arrow going up indicate addition, going down substraction. For the general
  $L$ dimensional hypercubes $L$ cycles are necessary where terms differing at
  different bits are combined.}
  \label{fig:FFT}
\end{center}
\end{figure}

An instance of \texttt{hypercube\_lowd} can be initialized with the function values of the hypercube or with
its Fourier coefficients. The population class, \texttt{haploid\_lowd}, holds instances of \texttt{hypercube\_lowd} for the population,
the mutant genotypes, the recombinant genotypes and the fitness function.

From a practical point of view, an instance of a low-dimensional population is initialized in three steps,
\begin{enumerate}[(a)]
 \item the class is instantiated
  \begin{lstlisting}
  haploid_lowd::haploid_lowd(int L=1, int rng_seed=0)
  \end{lstlisting}
  where L is the number of loci and \texttt{rng\_seed} is a seed for the random number generator (a random seed by default);
 \item the initial population structure is set by the functions
  \begin{lstlisting}
  int haploid_lowd::set_allele_frequencies(double *freq, unsigned long N)
  int haploid_lowd::set_genotypes(vector <index_value_pair_t> gt)
  int haploid_lowd::set_wildtype(unsigned long N)
  \end{lstlisting}
  set the population size and composition.
  The first function initializes random genotypes in linkage equilibrium with the specifiec allele frequencies \texttt{freq},
  the second explicitely sets a number of individuals for each genotype using the new type \texttt{index\_value\_pair},
  and the last one sets a wildtype-only population of size \texttt{N};
 \item the fitness hypercube is initialized directly by accessing the attribute
  \begin{lstlisting}
  haploid_lowd::fitness
  \end{lstlisting}
  in the population class.
\end{enumerate}

\subsubsection{Evolution}
In traditional Wright-Fisher type models, in each generation, the expected
frequencies of gametes with a particular genotype after mutation, selection, and
recombination  are calculated and then the population resampled from this
gamete distribution. We will now outline the steps required to update the
population. A more detailed discussion can be found in \citep{Neher:2011p45096}. 
All the following steps are called by either one of the following functions:
\begin{itemize}
\item \begin{lstlisting}
int haploid_lowd::evolve(int gen=1)
\end{lstlisting}
updates the population looping over a specified number of generations \texttt{gen};
\item \begin{lstlisting}
int haploid_lowd::evolve_norec(int gen=1)
\end{lstlisting}
is an alternative version that skips the resampling (deterministic evolution);
\item \begin{lstlisting}
int haploid_lowd::evolve_deterministic(int gen=1)
\end{lstlisting}
is another alternative that skips the recombination step (asexual evolution).
\end{itemize}

\subsubsection{Mutation}
Let $P(\gt)$ be the genotype distribution at the beginning of a generation.
Denoting the mutation rate towards the $1$ or $0$ state at
locus $i$ by $\mut_i^{1/0}$, the expected $P(\gt)$ after mutation will be
\begin{equation}
 P(\gt) \quad \longleftarrow \quad \left(1-\sum_{i=0}^{L-1} \mut_i^{\overline{s_i}} \right)P(\gt)
 + \sum_{i=0}^{L-1} \mut_i^{s_i} P(\Pi_i \gt)
\end{equation}
where $\Pi_i \gt$ denotes genotype $\gt$ with locus $i$ flipped from $0$ to $1$
or vice versa. The first term is the loss due to mutation, while the second term
is the gain due to mutation from neighbouring genotypes (in terms of Hamming distance).

The mutation rates can be specified by the folowing set of overloaded functions:
either a single double rate (same for every position and in both forward and backward sense), two double rates (forward and backward rates),
a $L$ dimensional double array (site-specific, identical forward and backward rates), or a $2\times L$ dimensional double array,
\begin{itemize}
 \item \begin{lstlisting}
int haploid_lowd::set_mutation_rate(double rate);
\end{lstlisting}
takes a single rate and sets it for every position and in both forward and backward sense;
\item \begin{lstlisting}
int haploid_lowd::set_mutation_rate(double rate_forward, double rate_backward);
\end{lstlisting}
takes two rates and sets \texttt{rate\_forward} as the forward rate ($0 \rightarrow 1$) and \texttt{rate\_backward} as the backward rate ($1 \rightarrow 0$)
\item \begin{lstlisting}
int haploid_lowd::set_mutation_rate(double* rates);
\end{lstlisting}
takes the pointer to an array of length L and sets site-specific rates, the same for forward and backward mutations;
\item \begin{lstlisting}
int haploid_lowd::set_mutation_rate(double **rates)
\end{lstlisting}
takes a pointer to a pair of (pointers to) arrays, each of length L, which contain the site-specific rates for forward (\texttt{rates[0]}) and backward (\texttt{rates[1]}) mutations.
\end{itemize}

\subsubsection{Selection}
Selection reweighs different the population of different genotypes according to
their fitness as follows
\begin{equation}
P(\gt) \quad \longleftarrow \quad \frac{e^{F(\gt)}}{\mexpfit} P(\gt)
\end{equation}
where $\mexpfit$ is the population average of $e^{F(\gt)}$, which is required to keep the population size constant.
The corresponding function is
\begin{lstlisting}
int haploid_lowd::select_gametes()
\end{lstlisting}

\subsubsection{Resampling and genetic drift}
For deterministic modeling, one generation would be completed at this point and
one would repeat the cycle, starting with mutation again. For stochastic population
genetics, we still need to resample the population in a way that mimics the
randomness of reproduction. The easiest and most generic way to do this is to
resample a population of size $N$ using a multinomial distribution with the
current $P(\gt)$ as sampling probabilities of different genotypes.
Alternatively, one can sample individuals according to a Poisson distribution
with mean $NP(\gt)$ for each genotype, which will result in a population of
approximately size $N\pm \mathcal{O}(\sqrt{N})$. For large populations, the two
ways of resampling are equivalent and we chose the latter (much faster)
alternative. The function
\begin{lstlisting}
int haploid_lowd::resample()
\end{lstlisting}
samples the next generation the expected genotype frequencies. The expected
population size used in the resampling is the carrying capacity.

\subsubsection{Mating and recombination}
The computationally expensive part of the dynamics is recombination, which needs
to consider all possible pairs of pairs of parents and all different ways in
which their genetic information can be combined.
In a facultatively sexual population, a fraction $\ox$ of the individuals undergo
mating and recombination. In obligate sexual populations, $\ox=1$.
The genotype distribution is updated according to the following rule:
\begin{equation}
 P(\gt) \quad \longleftarrow \quad \left(1 - \ox\right)P(\gt)
 + \ox R(\gt)
\end{equation}

The distribution $R(\gt)$ of recombinant gametes would naively be computed as follows:
\begin{equation}
R(\gt) = \sum_{ \xi } \sum_{\gt'} C(\xi) P(\gt^m)P(\gt^p),
\end{equation}
where $\xi$ specifies the particular way the parental genomes are combined:
$\xi_i=0~\text{(resp. $1$)}$ if locus $i$ is derived from the mother (resp. father).
The genotype $\gt'$ is summed over; it represents the part of the maternal
($\gt^{m}$) and paternal ($\gt^{p}$) genotypes that is not passed on to the
offspring.
We can decompose each parent into successful loci that made it into the
offspring and wasted loci, as follows: $\gt^{p} = \xi \land \gt + \overline{\xi}
\land \gt'$ and $\gt^{m} = \overline{\xi} \land \gt + \xi \land \gt'$, where
$\land$ and a bar over a variable indicate respectively the elementwise AND and NOT operators
(i.e., $\overline{\xi_i} := 1 - \xi_i$). The function $C$ assigns a probability
to each inheritance pattern. Depending on whether the entire population
undergoes sexual reproduction or only a fraction $\ox$ of it, the entire
population or a fraction $\ox$ is replaced with $R(\gt)$.
The central ingredient for the efficient computation of $R(\gt)$ is the Fourier
decomposition introduced above. The generic Fourier coefficient of $R(\gt)$ is
given by
\begin{equation}
 r^{(k)}_{i_1\ldots i_k} = 2^{-L} \sum_{\gt} \locuspm_{i_1}\ldots \locuspm_{i_k} \left(
 \sum_{\xi} \sum_{\gt'}
C(\xi) P(\gt^m)P(\gt^p) \right) \\
\end{equation}
Just as $\gt^{p}$ and $\gt^{m}$ can be expressed as a combination of $\gt$ and
$\gt'$, we can invert the relation and express the generic $t_i$ as a function
of $\gt^{p}$ and $\gt^{m}$, as follows: $t_i =
\xi_i\locuspm^m_i+\overline{\xi_i}\, \locuspm^p_i$. Using this new basis and
exchanging the order of summations, we obtain
\begin{equation}
 r^{(k)}_{i_1\ldots i_k} = 2^{-L} \sum_{\xi}
C(\xi)
 \sum_{\gt^m, \gt^p}
(\xi_{i_1}\locuspm^m_{i_1}+\overline{\xi_{i_1}}\, \locuspm^p_{i_1})\ldots
(\xi_{i_k}\locuspm^m_{i_k}+\overline{\xi_{i_k}}\, \locuspm^p_{i_k})
P(\gt^m)P(\gt^p).
\end{equation}
Notice that $C(\xi)$ can be pulled out of the two inner sums, because the odds
of inheriting a certain locus by the mother/father is independent of what their
genetic makeup looks like.
Next we expand the product and introduce new labels for compactness,
\begin{equation}
 r^{(k)}_{i_1\ldots i_k} = 2^{-L} \sum_{\xi}
 C(\xi) 
\sum_{\gt^m, \gt^p}
P(\gt^m)P(\gt^p)
\sum_{l=0}^k \sum_{\{j_i\}, \{h_i\}}
\xi_{j_1}\ldots\xi_{j_l}\overline{\xi_{h_1}}\ldots\overline{\xi_{h_{k-l}}} \, \locuspm^m_{j_1}\ldots\locuspm^m_{j_l}
\locuspm^p_{h_1}\ldots\locuspm^p_{h_{k-l}},
\end{equation}
where $l$ is the number of loci inherited from the mother among the $k$ in $(i_1,\ldots, i_k)$.
$l$ runs from $0$ (everything happens to be contributed by the father)
to $k$ (everything from the mother). $\{j_i\}$ and $\{h_i\}$ are all (unordered) partitions of $i$ into sets of size $l$ and $k-l$, respectively.
Now we can group all $\xi_i$ in the inner sum with $C(\xi)$, all $t_i^{m}$ with $P(\gt^m)$, and all $t_i^{p}$ with $P(\gt^p)$.
The three sums (over $\xi$, $\gt^m$, and $\gt^p$) are now completely decoupled. Moreover, the two sums over the parental genotypes
happen to be the Fourier decomposition of $P(\gt)$. Hence, we have
\begin{equation}
 r^{(k)}_{i_1\ldots i_k} = \sum_{l=0}^k \sum_{\{j_i\}, \{h_i\}} C^{(k)}_{j_1\ldots
j_l, h_1\ldots h_{k-l}} p^{(k)}_{j_1\ldots j_l} p^{(k-l)}_{h_1\ldots h_{k-l}}.
\end{equation}
The quantity 
\begin{equation}\label{eq:reccoeff}
C^{(k)}_{j_1\ldots j_l,h_1\ldots h_{k-l}} = \sum_{\xi}
C(\xi)\xi_{j_1}\ldots\xi_{j_l}\overline{\xi_{h_1}}\ldots\overline{\xi_{h_{k-l}}}
\end{equation}
can be calculated efficiently, for each pair of partitions $(\{j_i\}, \{h_i\})$, by realizing that (a) for $k=L$,
there is exactly one term in the sum on the right that is non-zero and (b) all
lower-order terms can be calculated by successive marginalizations over unobserved
loci. For instance, let us assume that $k=L-1$ and that the only missing locus is the m-th one.
We can compute
\begin{equation}
C^{(L-1)}_{j_1\ldots j_l,h_1\ldots h_{L-1-l}} =  \;
C^{(L)}_{j_1\ldots j_l \, m,h_1\ldots h_{L-1-l}} + \; 
C^{(L)}_{j_1\ldots j_l,h_1\ldots h_{L-1-l}\, m}.
\end{equation}
There are $\binom{L}{k}$ ways of choosing $k$ loci out of $L$, which can
be inherited in $2^k$ different ways (the partitions in $j$ and
$h$ in \EQ{reccoeff}) such that the total number of coefficients is $3^L$.
Note that these coefficients are only calculated when the recombination rates change.
Furthermore, this can be done for completely
arbitrary recombination patterns, not necessarily only those with independent
recombination events at different loci. 

\texttt{haploid\_lowd} provides a function to calculate $C(\xi)$ from
recombination rates between loci assuming a circular or linear chromosome. 
The probability of a particular crossover pattern is calculated assuming
independent crossovers. The function 
\begin{lstlisting}
int haploid_lowd::set_recombination_rate(double *rec_rates)
\end{lstlisting}
assumes a double array of length $L-1$ for a linear chromosome and of length $L$
for a circular chromosome. For a linear (resp. circular) chromosome, the i-th element
of the array is the probability of recombining after (resp. before) the i-th locus.
Furthermore, the mating probability $\ox$ must be specified explicitely via the attribute
\begin{lstlisting}
haploid_lowd::outcrossing_rate
\end{lstlisting}
the default is obligate sexual reproduction.

The code offers a simpler alternative for free recombination. In this case, only the global
mating probability $\ox$ needs to be entered. If the user does not set the recombination rates
via \texttt{set\_recombination\_rate}, free recombination is the default behaviour. Otherwise,
this option is controlled by the following boolean attribute
\begin{lstlisting}
haploid_lowd::free_recombination
\end{lstlisting}

Note that, in a circular chromosome, there is effectively one more inter-locus
segment (between the last and the first locus) in which crossovers can occur,
and the total number of crossovers has to be even. Assuming independent
crossovers, the global recombination rate of circular chromosomes is lower than
a linear chromosome of the same length by a factor of
$(1-e^{-r_0})$, where $r_0$ is the recombination rate between the
first and last loci.

The recombination process itself is initiated by 
\begin{lstlisting}
int haploid_lowd::recombine()
\end{lstlisting}

\subsection{FFPopSim for many loci}
For more than $20$ loci, storing then entire genotype space and all possible
recombinants becomes prohibitive. Hence we also include a streamlined individual
based simulation package that can simulate arbitrarily
large number of loci and has an overall runtime and memory requirements
$\mathcal{O}(NL)$ in the worst case scenario. The many-loci package uses the same interface as
the few-loci one. This makes it easy, for example, to first test an evolutionary scenario using
many (all) loci and to focus on the few crucial ones afterwards.

To speed up the program in many cases of interest, identical genotypes are grouped into clones.
This part of the library was developed for whole genome simulations of large HIV populations
($L=10^4$, $N>10^5$). A specific wrapper class for HIV applications is provided.

\subsubsection{Specification of the population and the fitness func}
For more than 20 loci, it becomes infeasible to store the entire hypercube.
Instead, we store individual genotypes as bitsets. Each genotype, together with
the number of individuals that carry it, as well as traits and fitness
associated with it is stored for as long as it is present in the population. All of
this is aggregated in the structure \texttt{clone\_t}. The population is a vector of
clones. Each generation clones size are updated and added to the new generation,
new clones are produced, and empty ones deleted.

Fitness functions are again functions on the hypercube. The latter is implemented as
\texttt{hypercube\_highd}. Instead of storing all possible fitness values, 
\texttt{hypercube\_highd} stores non-zero Fourier coefficients. Whenever a new
genotype is produced, its fitness is calculated by summing the appropriate
coefficients.

\subsubsection{Mutation}
To implement mutation, a Poisson distributed number $n$ with mean $N\mu_i$ is
drawn for each locus $i$ and $n$ mutations are introduced at locus $i$ into $n$
randomly chosen genotypes. Mutations are bit-flip operations in the bitset.
Only a global mutation rate is currently supported.

\subsubsection{Selection}
Prior to selection, the population average $\overline{W}:=\mexpfit$
and a growth rate adjustment $\exp(1-N/N_0)$ are computed. The latter is used to
keep the population size close to the carrying capacity $N_0$. The size of each
clone is then updated with a Poisson distributed number with mean
$(1-r)\overline{W}^{-1}\exp(1-N/N_0)$, where $r$ is the recombination rate.
Another Poisson distributed number with mean $r\overline{W}^{-1}\exp(1-N/N_0)$ is
set aside for recombination later.

\subsubsection{Recombination}
The individuals marked for sexual reproduction during the selection step are
shuffled and paired. For each pair, a bitset representing the crossover pattern
is produced and two new clones are produced from the two parental genomes.
Alternatively, all loci of the two genomes can be reassorted at random.

\subsection{Python wrapper}
The C++ library includes Python bindings that greatly simplify interactive use and testing.
The wrapping itself is done by SWIG~\citep{Beazley:2003:ASS:860016.860018}.
Most notably, the C++ classes \texttt{haploid\_lowd}, \texttt{haploid\_highd} and the
HIV-specific subclass are fully exposed to Python, including all their public members.
The performance speed for evolving a population is unchanged, since the \texttt{evolve}
function iterates all steps internally for an arbitary number of generations.

The bindings are not completely faithful to the C++ interface, to ensure a more intuitive
user experience. For instance, C++ attribute set/get members are translated into Python
properties via the builtin \texttt{property} construct.
Furthermore, since direct access to the \texttt{hypercube\_lowd} instances from Python
is not straightforward, a few utility functions have been written to do common tasks.
The fitness hypercube can be set easily by either one of
\lstset{language=Python, basicstyle=\small, showstringspaces=false, backgroundcolor=\color{white}, rulecolor=\color{black}, tabsize=2, keywordstyle=\color{mauve}, commentstyle=\color{dkgreen}, stringstyle=\color{blue} }
\begin{lstlisting}
haploid_lowd.set_fitness_function(genotypes, fitnesses)
haploid_lowd.set_fitness_additive(fitness_main)
\end{lstlisting}
The former function is used to set specific points on the hypercube:
it takes a sequence of genotypes \texttt{genotypes} (integers or binary literals using \texttt{0b} notation)
and a sequence of fitness values \texttt{fitnesses}, corresponding to those genotypes. Any missing
point on the fitness hypercube will be consider neutral. The second function creates an additive fitness landscape,
in which main effects are specified by the L-dimensional input sequence \texttt{fitness\_main}.

After installation, the FFPopSim library can be used in Python as a module, e.g.
\begin{lstlisting}
from FFPopSim import haploid_lowd
\end{lstlisting}
The bindings make heavy use of the NumPy library and its SWIG fragments and
typemaps~\citep{oliphant2006guide}.
We therefore recommend to import NumPy before FFPopSim, although this is not strictly necessary.
Moreover, the Python binsings include a few functions for plotting features of the population, such as
genetic diversity. The Python module Matplotlib is required for this purpose~\citep{2878517}.

The HIV-specific part of the code has been expanded further in Python to enable quick simulations of viral evolution
under commonly studied conditions. In particular, random genotype-phenotype maps for viral replication capacity
and drug resistance can be generated automatically from a few parameters, via the functions
\begin{lstlisting}
hivpopulation.set_replication_landscape
hivpopulation.set_resistance_landscape
\end{lstlisting}
The input parameters reflect a number of typical properties of HIV populations, such as the fraction of sites
carrying potentially adaptive mutations. See the inline Python documentation for further details on these functions.
Moreover, since studies of HIV evolution often involve a large number of genotypic data, a function for saving the 
genotype of random individuals from the current population in a compressed format has been added. The syntax is the following:
\begin{lstlisting}
hivpopulation.write_genotypes_compressed(filename, number_of_individuals)
\end{lstlisting}
where \texttt{filename} is the name of the file, in which the data are to be stored, and \texttt{number\_of\_individuals} if the
size of the random sample. The data can be accessed again by the standard Numpy \texttt{load} function.

\clearpage

\section{Examples}
\subsection{\texttt{haploid\_lowd}: Valley crossing}
One of the most striking effects of genetic epistasis is the slowdown of evolution when a combination of mutations is
beneficial, but intermediate mutants are deleterious compared to wildtype.
Such scenario is relevant in applications, for instance for the emergence of bacterial or viral resistance
to drugs~\citep{Weinreich:2005p15528}.
Not surprisingly, recombination plays a central role in this process. On the one hand, it enhances the production rate
of multiple mutants, on the other it depletes the class of complete mutants by back-recombination with deleterious
backgrounds.

FFPopSim makes the efficient simulation of such processes as easy as the
following script:
\begin{lstlisting}
import FFPopSim
L = 4 # number of loci
N = 1e10 # population size
s1=1e-5 # fitness of wildtype
s2=0.01 # fitness of quadruple mutant

c = FFPopSim.haploid_lowd(L) # create population
c.set_genotypes([0b0],[N]) # start with wildtype

c.set_recombination_rates([1e-2] * (L-1)) # set recombination rates
c.set_mutation_rate(1e-5) # set mutation rate

# assign positive relative fitness to wildtype and quadruple mutant
c.set_fitness_function([0b0, 0b1111], [s1, s1+s2])

# cross valley with an accuracy of 100 generations
gens_at_once = 100
while c.get_genotype_frequency(0b1111)<0.5 and c.generation<1e7:
    c.evolve(gens_at_once)

# print result
print 'Time to cross the valley: '+str(c.generation)+' generations'
\end{lstlisting}

\begin{figure}[htp]
\begin{center}
  \includegraphics[width=0.5\columnwidth]{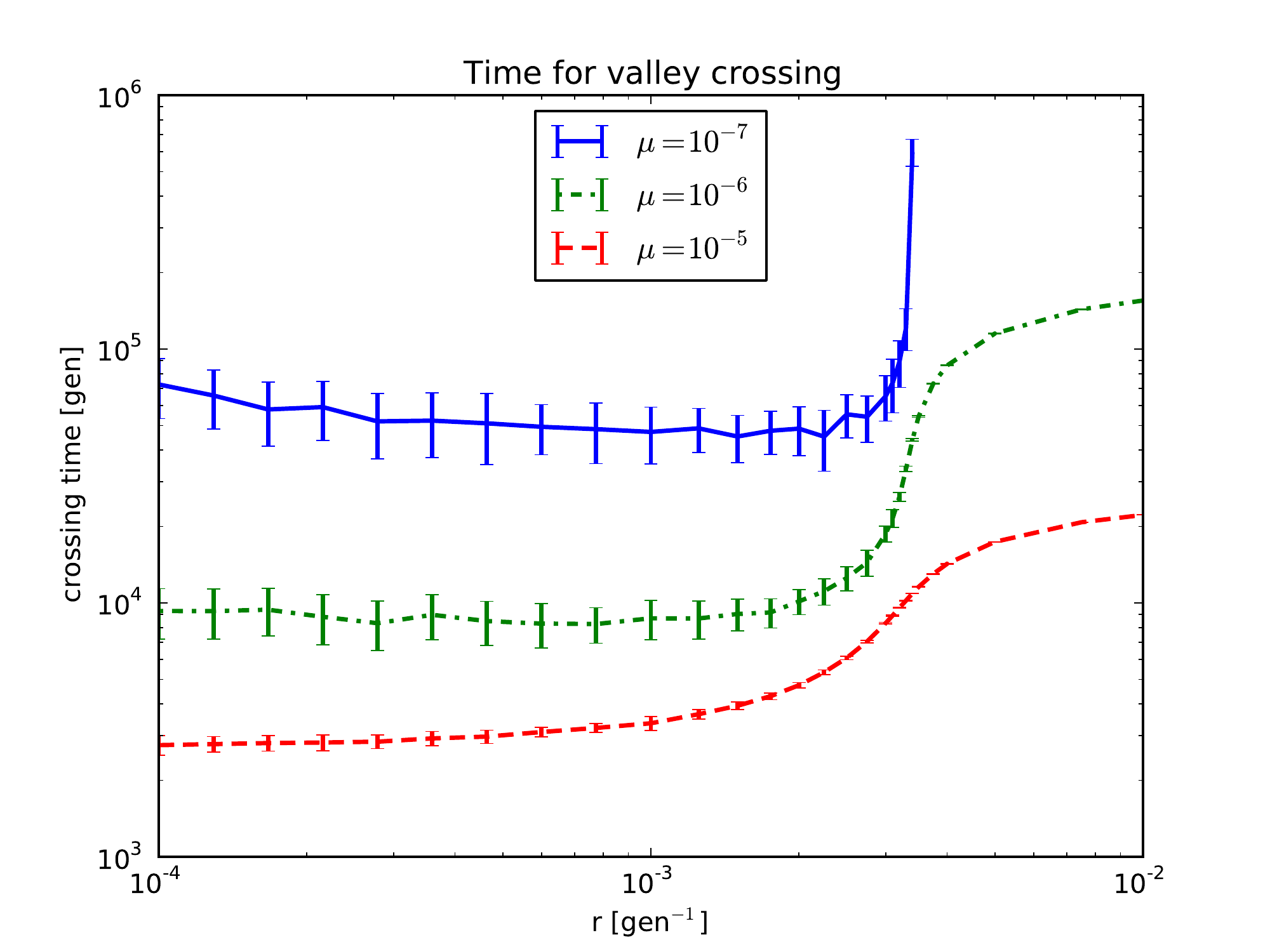}
  \caption[labelInTOC]{Example of the Python wrapper for \texttt{haploid\_lowd}.
  The average time for crossing a 4 dimensional fitness
  valley is plotted against the recombination rate. Bars indicate the standard
  deviation across fifty repetitions. Crossing a valley becomes essentially
  impossible once the total recombination rate exceeds the fitness benefit,
  at least when mutation rates are small.}
  \label{fig:valley}
\end{center}
\end{figure}

If the script is run with different recombination rates, the effect of this parameter on the time for valley crossing can
be investigated, as shown in \FIG{valley}. The full scripts producing the
figures is provides as separate file.
\clearpage

\subsection{\texttt{haploid\_lowd}: HIV immune escape}
During an HIV infection, the host immune system targets several viral epitopes
simultaneously via a diverse arsenal of cytotoxic T-cells (CTLs).
Mutations at several loci are thus selected for and start to rise in frequency at the same time but,
because of the limited amount of recombination, end up in wasteful competition (interference) at frequencies of order one.

\begin{figure}[htp]
\begin{center}
  \includegraphics[width=0.5\columnwidth]{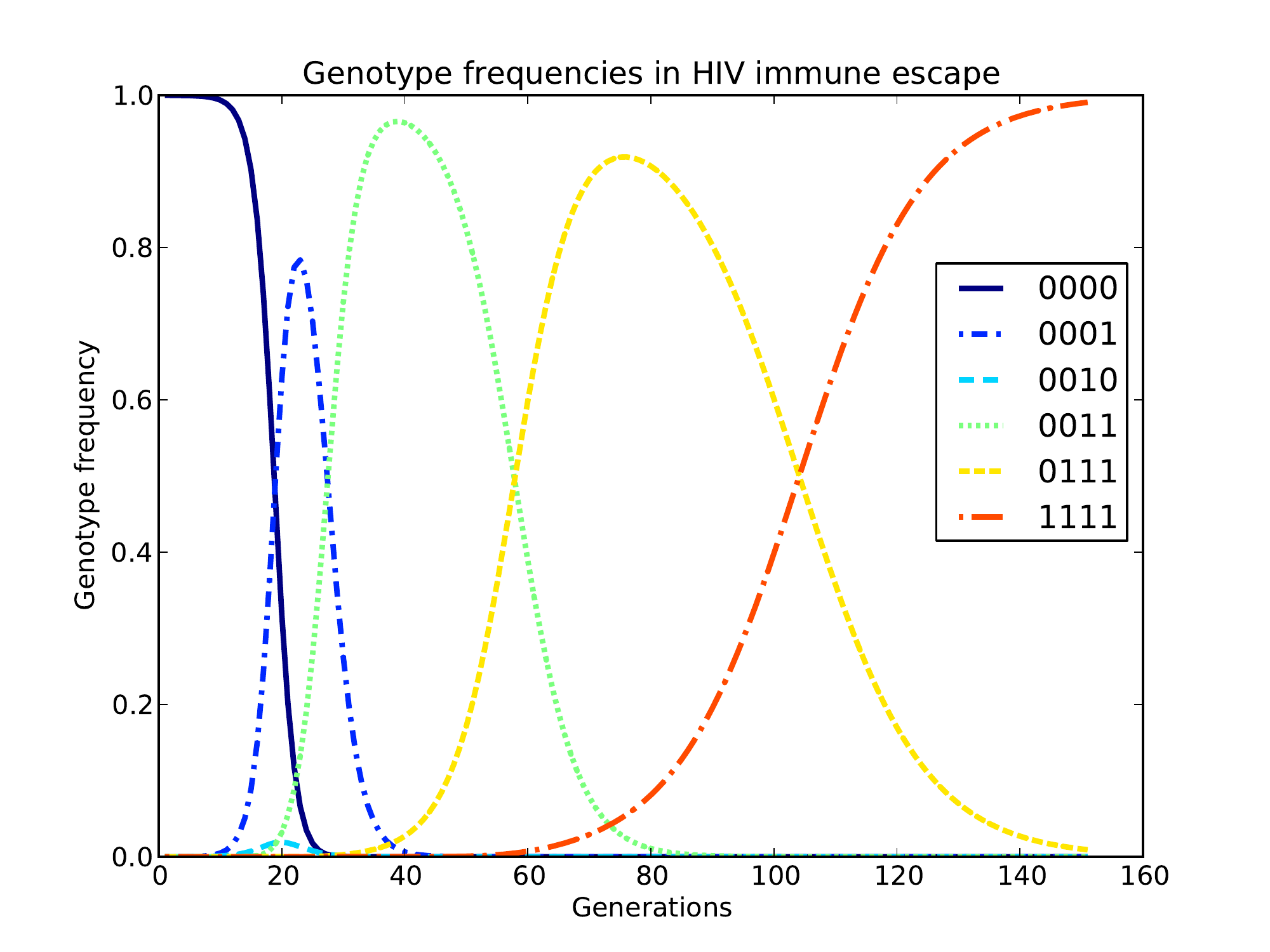}
  \caption[labelInTOC]{Example of the Python wrapper for \texttt{haploid\_lowd}. The frequencies of some genotypes in the population during
  strong, parallel positive selection are shown.}
  \label{fig:HIVescape}
\end{center}
\end{figure}

The theoretical description of genetic interference is involved and often limited to two-loci models,
but FFPopSim makes the simulation of this process straightforward. The following script evolves a 4-loci population under
parallel positive selection and tracks its genotype frequencies:
\begin{lstlisting}
import FFPopSim
L = 4 # number of loci
N = 1e10 # population size

c = FFPopSim.haploid_lowd(L) # create population
c.set_genotypes([0b0],[N]) # start with wildtype

c.set_recombination_rates([1e-4] * (L-1)) # set recombination rates
c.set_mutation_rate(1e-5) # set mutation rate

pop.set_fitness_additive([0.3, 0.2, 0.1, 0.05]) # set an additive fitness landscape

# evolve until fixation of the quadruple mutant,
# storing times and genotype frequencies
times = []
genotype_frequencies = []
while pop.get_genotype_frequency(0b1111) < 0.99 and pop.generation<1e7:
    pop.evolve()
    times.append(pop.generation)
    genotype_frequencies.append(pop.get_genotype_frequencies())
\end{lstlisting}
The resulting genotype frequencies are shown in \FIG{HIVescape}. The full scripts producing the
figures is provides as separate file.
\clearpage

\subsection{\texttt{haploid\_highd}: HIV chronic infection}
HIV evolution during chronic infection is determined by a number of parallel processes, such as mutation, recombination,
and selection imposed by the immune system. In combination, these processes
give rise to a complicated dynamics and don't understand how population
features such as population diversity depend on model parameters. Hence
simulations are an important source of insight. 

\begin{figure}[htp]
\begin{center}
  \includegraphics[width=0.5\columnwidth]{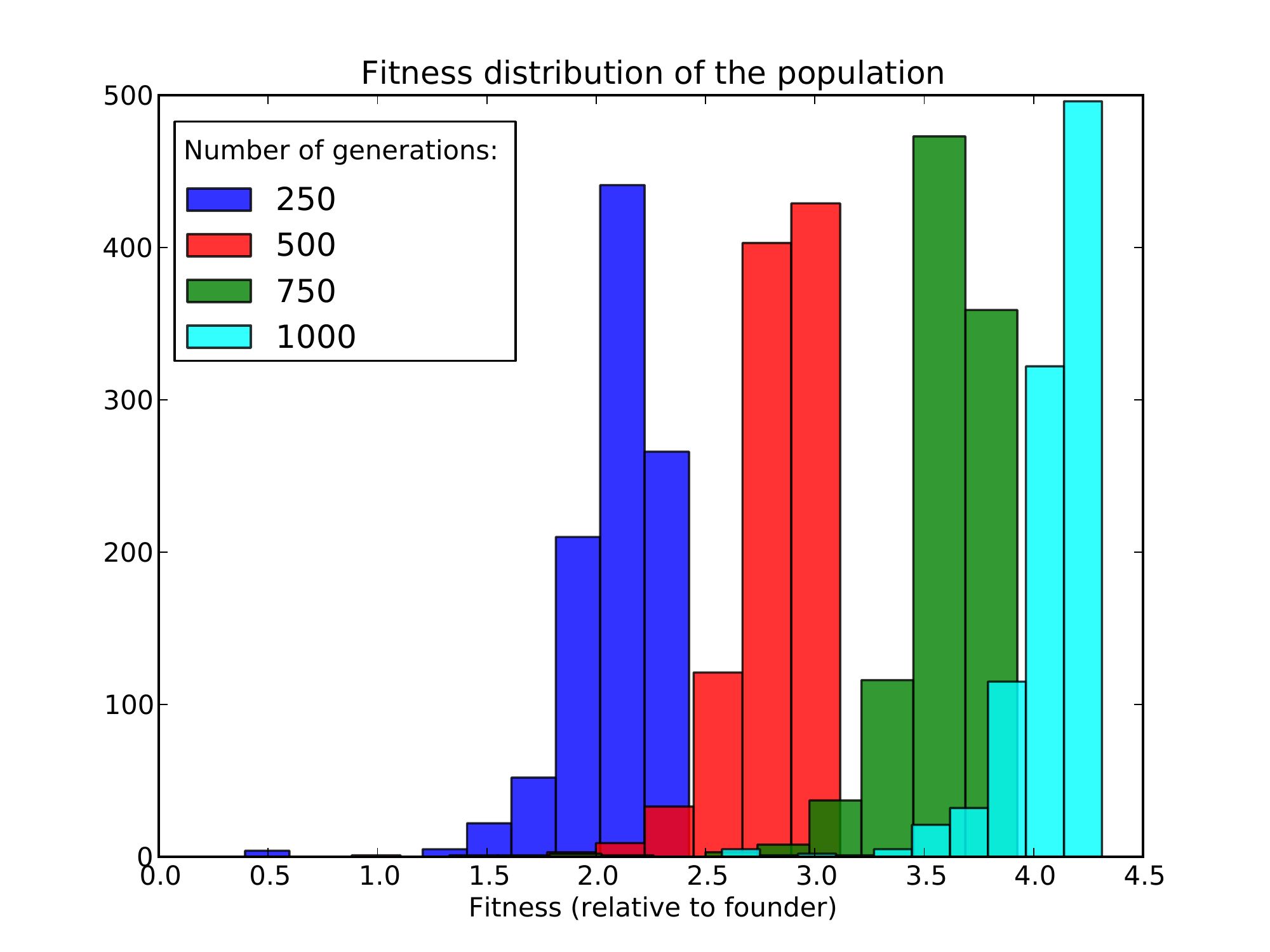}
  \caption[labelInTOC]{Example of the Python wrapper for \texttt{hivpopulation}. The fitness distribution of the population
  after 250, 500, 750, and 1000 generations from transmission is shown.}
  \label{fig:HIVfitness}
\end{center}
\end{figure}

FFPopSim offers a specific double C++/Python interface to this problem via its class \texttt{hivpopulation}.
The following script simulates an HIV population for one thousand generations, under a random static fitness landscape,
and stores a hundred random genomes from the final population in a compressed file:
\begin{lstlisting}
import FFPopSim
N = 1000 # population size

# create the population. Default options:
# - wildtype only
# - no treatment, i.e. replication is the same as fitness
pop = FFPopSim.hivpopulation(N)

# set a random, additive replication capacity landscape using
# the following parameters
pop.set_replication_landscape(adaptive_fraction=0.01,
                              effect_size_adaptive=0.03)

# evolve the HIV population
pop.evolve(1000)

# store final genomes in compressed format for further analysis
pop.write_genotypes_compressed('HIV_genomes.npz', 100)
\end{lstlisting}
NumPy can be used subsequently to analyze the genome sequences. Alternatively, the internal Python functions can be used,
e.g. for calculating the fitness distribution directly using
\texttt{hivpopulation.get\_fitness\_histogram}, as shown in \FIG{HIVfitness}. The full scripts producing the
figures is provides as separate file.
\clearpage


\end{document}